\documentclass[lettersize,journal]{IEEEtran}

\usepackage{tikz}

\usepackage{graphicx}
\usepackage[lined,ruled,commentsnumbered]{algorithm2e}
\usepackage{cite}
\usepackage[numbers,sort&compress]{}
\usepackage{amsmath,amssymb,amsfonts}
\usepackage{array}
\usepackage{graphicx}
\usepackage{makecell} 
\usepackage{amsthm}

\newtheorem{definition}{Definition}
\usepackage{threeparttable}  

\usepackage{textcomp}
\usepackage{xcolor}
\usepackage{booktabs}
\usepackage{multirow}
\usepackage{subfigure}
\usepackage{longtable}
\usepackage{threeparttable}
\usepackage{enumitem}
\usepackage{amssymb}
\usepackage{soul}
\usepackage{algorithmic}
\usepackage{textcomp}
\usepackage{xcolor}
\usepackage{array}
\usepackage{url}
\usepackage{float}
\usepackage{bm}
\usepackage{adjustbox}
\usepackage{epstopdf}

\usepackage{subcaption}
\usepackage{diagbox}

\usepackage{amsthm}
\usepackage{tabularx}
\usepackage{mathrsfs} 
\usepackage{soul}

\theoremstyle{remark}

\theoremstyle{plain} 

\usepackage{stfloats}
\usepackage[colorlinks=true, linkcolor=blue, citecolor=blue]{hyperref}

\usepackage{cleveref}
\begin{document}

\def\BibTeX{{\rm B\kern-.05em{\sc i\kern-.025em b}\kern-.08em
    T\kern-.1667em\lower.7ex\hbox{E}\kern-.125emX}}


\title{Harnessing Environmental Scatterers for Endogenous Target Covertness in 6G ISAC: A New Paradigm for Sensing Security}

\author{Yifan~Zhang,~\IEEEmembership{Member,~IEEE,}
        Boxuan~Xie,~\IEEEmembership{Member,~IEEE,}
        Mingyang~Yang,
        Ying~Liu,~\IEEEmembership{Student Member,~IEEE,}
        Riku Jäntti,~\IEEEmembership{Senior Member,~IEEE,}
        Zan~Li,~\IEEEmembership{Fellow,~IEEE.} 


\thanks{Y. Zhang B. Xie, Y.~Liu, and R. Jäntti are with the Department of Information and Communications Engineering, Aalto University, Espoo, 02150, Finland. (email: yifan.1.zhang@aalto.fi, boxuan.xie@aalto.fi,  ying.2.liu@aalto.fi, riku.jantti@aalto.fi)}
\thanks{Mingyang~Yang and Z. Li (Corresponding Author) are with the State Key Laboratory of Integrated
Services Networks, School of Telecommunication Engineering,
Xidian University, Xi’an 710071, China (email: zanli@xidian.edu.cn, yangmingyang@ceyear.com)}
 }

\markboth{}%
{Shell \MakeLowercase{\textit{et al.}}: A Sample Article Using IEEEtran.cls for IEEE Journals}

\maketitle

\begin{abstract}
Integrated sensing and communication (ISAC) is a cornerstone of 6G networks, enabling simultaneous data transmission and environmental perception. However, the ubiquitous radiation of sensing waveforms introduces a critical security vulnerability where the physical existence and characteristics of targets are continuously exposed to unauthorized wardens. Meanwhile, existing sensing-secure methods primarily focus on degrading a target warden's parameter-estimation accuracy, yet they fail to prevent wardens from detecting the target's physical presence. In this article, we first analyze sensing security threats in ISAC systems and survey existing sensing-secure methods. To address the critical gap, we propose a novel security paradigm termed endogenous target covertness. By intentionally harnessing environmental scatterers, such as known objects, ambient backscatter devices, and reconfigurable intelligent surfaces (RIS), as ghost targets, the proposed scheme ensures a low probability that the warden detects the true target for the true target at the warden. This deliberately renders the true target's existence indistinguishable to the warden, while the legitimate base station can extract the true target's response using prior knowledge of the scatterers. We demonstrate the effectiveness of the proposed scheme through a case study, proving that absolute target covertness can be achieved without consuming dedicated jamming power. Finally, we outline key open challenges and promising research directions to realize truly secure, ubiquitous sensing in future ISAC ecosystems.
\end{abstract}





\section{Introduction}



Integrated sensing and communication (ISAC) is widely recognized as one of the key enabling technologies for sixth-generation (6G) wireless networks \cite{10418473}. Using shared hardware architectures, spectrum resources, and signal processing frameworks, ISAC enables the simultaneous execution of data transmission and environmental perception. This profound convergence not only improves spectral and energy efficiency, but also provides ubiquitous target awareness, which can facilitate many emerging 6G applications, including autonomous driving, immersive virtual reality (VR), and the burgeoning low-altitude economy ecosystem \cite{10216343,9737357}. More fundamentally, ISAC bridges the digital and physical worlds and establishes a new foundation for context-aware, autonomous, and intelligent operation of future 6G networks.

However, this paradigm shift introduces unique security issues in sensing. Since sensing waveforms and target-induced echoes inherently radiate into open wireless propagation environments, the physical presence, location, and motion of targets are continuously exposed in the surrounding space \cite{9755276}. This open nature enables malicious target wardens to intercept echo signals and infer highly sensitive physical information \cite{11202391}. Unlike traditional communication scenarios where data privacy is reliably guaranteed by sophisticated cryptographic protocols \cite{10044183}, the physical characteristics and existence of a target are deeply embedded in unencrypted analog sensing signals. This inherent physical exposure bypasses traditional cybersecurity defenses, creating a critical security gap within current ISAC frameworks.

More recently, many schemes have made preliminary contribution to mitigate these risks by employing artificial noise (AN) to reduce the detection probabilities of malicious wardens \cite{10587082}, utilizing advanced beamforming techniques to minimize the Cramér-Rao bound (CRB) of legitimate sensing receivers while degrading target warden channels \cite{10804654}, and applying ambiguity function engineering to transmit signals that synthesize false targets \cite{11202391}. Nevertheless, a fundamental limitation persists. These approaches are primarily designed to degrade the parameter estimation accuracy of a warden after the target has already been observed. They fail to prevent the target warden from detecting the mere existence of the target. Once a target's physical presence is exposed, malicious entities can trivially infer sensitive behavioral routines, initiate unauthorized localized tracking, or launch targeted cyber-physical attacks. Consequently, there is a pressing need to shift from simply degrading target warden accuracy to developing an effective security paradigm capable of completely concealing a target's existence.
\begin{figure}[tp]
    \centering
    \includegraphics[width=0.9\linewidth]{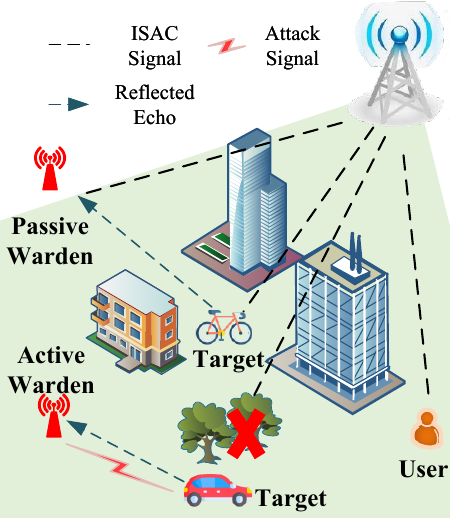}
    \caption{Target Warden in ISAC Systems.}
    \label{attack}
\end{figure}

Motivated by this gap, this article advocates target covertness as a new sensing-security paradigm for future ISAC systems. We first review emerging sensing-security threats and representative countermeasures. We then introduce target covertness to conceal the physical existence of targets from unauthorized wardens. By leveraging nearby scatterers \cite{115famaey2026survey75589}, including background objects, ambient backscatter devices, and reconfigurable intelligent surfaces (RIS), as fake targets, the proposed approach creates endogenous uncertainty around the true target. Numerical results are presented to validate its effectiveness, followed by a discussion of key open issues and future research directions. Specifically, this article provides the following three perspectives:
\begin{enumerate}
    \item 
    We first provide a systematic overview of the emerging sensing security threats to 
    ISAC sensing and categorize representative countermeasures, including secure 
    beamforming, AN-assisted sensing protection, and sensing-secure waveform design. Based on this review, we 
    identify the distinction between degrading unauthorized sensing accuracy and target covertness.

    \item 
    We formulate the concept of target covertness and illustrate how environmental 
    scatterers can be harnessed to realize LPTD in ISAC. We first formally define target covertness and propose an endogenous covert target scheme using known scatterers around the target as ghost targets. Then, numerical experiments are presented to demonstrate the effectiveness of the proposed scheme.

    \item 
    Finally, we discuss the key challenges toward practical target-covertness 
    systems. We identify current open issues and suggests future research directions. These 
    challenges reveal several opportunities for developing sensing security from 
    conventional adversarial suppression toward environment-native and endogenous 
    protection mechanisms for future 6G ISAC ecosystems.
\end{enumerate}


\section{Sensing Security in ISAC: Threats and Existing Defenses} \label{sec:background}

\subsection{Target Wardens in ISAC Systems}
As shown in Fig. \ref{attack}, a target warden represents a malicious adversary dedicated to illicitly detecting the physical presence or extracting sensitive parameters of a legitimate target in ISAC systems. Depending on their operational capabilities and attack strategies, these adversaries generally manifest in two distinct forms: passive and active wardens.
\subsubsection{Passive Warden}
The passive warden operates as a silent observer in the electromagnetic environment, aiming to breach confidentiality without transmitting signals. By covertly capturing the sensing signals emitted by the legitimate ISAC transmitter and their subsequent reflections off surrounding objects, this attacker analyzes the multipath signatures to infer highly sensitive physical information. Without triggering any security alarms, the passive attacker can deduce the architectural layout of a room, track the trajectories of moving targets, or even extract fine-grained biometric data such as human gestures, breathing patterns, and heartbeats.

\subsubsection{Active Warden}
In contrast to mere silent observation, the active warden aggressively participates in the electromagnetic exchange to extract sensing information or manipulate channel estimations. This adversary intentionally transmits deceptive pilot signals or unauthorized probing waveforms to the ISAC nodes to trick the system into revealing its beamforming parameters or target-tracking configurations. By actively forcing the legitimate transmitter to adjust its directional beam towards the target warden, or by independently illuminating a target to capture the resulting backscatter, the active attacker creates favorable conditions to extract precise spatial and sensing data that would otherwise remain securely hidden.

\subsection{Existing Defenses for Sensing Security}

\subsubsection{Artificial-Noise-Assisted Protection}
AN is a physical-layer security technique that transmits jamming signals alongside the intended ISAC waveforms. Its core principle is to project carefully designed noise toward unauthorized wardens, thereby degrading their received signal-to-noise ratio (SNR) and parameter estimation accuracy. Zou \textit{et al.}~\cite{10587082} demonstrated the effectiveness of AN for securing sensing functionality in ISAC networks. A key advantage is its ability to directly mask target echoes without requiring complex encryption algorithms. However, allocating power to AN reduces the energy available for legitimate sensing and communication, lowering overall energy efficiency. Moreover, imperfect or time-varying warden channel state information (CSI) may cause AN leakage into legitimate sensing directions, resulting in self-interference and degraded sensing performance.

\subsubsection{Secure Beamforming}
Beamforming design leverages multiple-input multiple-output (MIMO) antenna arrays to spatially focus transmitted energy towards legitimate targets or communication users while forming deep nulls in the directions of suspected wardens. Recent studies have extensively explored this approach. For instance, Bozanis \textit{et al.} optimized secure beamforming to minimize the Bayesian Cramér-Rao Bound (BCRB) under target-angle uncertainty \cite{11602055}, while Ren \textit{et al.} expanded this approach to secure cell-free ISAC networks against simultaneous information and sensing wardens \cite{10605793}. 
Jia et al.~\cite{10804654} proposed an illegal sensing suppression framework that employs transmit beamforming to impair both unauthorized target detection and direction-of-arrival estimation by maximizing the adversary's missed-detection probability or Cramér--Rao bound. 
 The primary advantage of secure beamforming is its ability to concurrently optimize legitimate sensing accuracy and communication throughput without the extra power overhead required by AN. Nevertheless, its primary limitation is that the efficacy of spatial nulling is profoundly dependent on the availability of highly accurate angle or location information regarding the target warden. In highly dynamic scenarios, or when dealing with passive and wardens with unknown locations or channels, the spatial degrees of freedom may be misaligned, rendering beamforming defenses significantly less reliable.

\subsubsection{Sensing-Secure Waveform Design}
Waveform design enhances sensing security by manipulating the time-frequency structure of ISAC signals to distort the resulting echoes and induce degraded, ambiguous, or false target parameters at the warden. For example, ambiguity-function engineering has been used to impair unauthorized sensing \cite{11202391}, while Wang \textit{et al.}~\cite{11004012} proposed a diffusion-model-based framework that embeds safeguarding signals into pilots to mask activity-induced CSI variations. A key advantage of waveform-level protection is that it does not necessarily require additional spatial degrees of freedom or large antenna arrays, and can even create decoy ``ghost'' targets. However, secure waveform synthesis increases computational complexity and may constrain communication performance. More importantly, these methods mainly distort target parameters rather than conceal the echo energy itself, and thus the target's physical existence may remain detectable.

\begin{table}[t]
    \centering
    \resizebox{\columnwidth}{!}{%
    \begin{threeparttable}
    \caption{Comparison with Existing Sensing Security Schemes}
    \label{table_1}

    \begin{tabular}{c|ccccc}
    \toprule[1.5pt]
    Works
    & Method
    & Covertness
    & Param. Prot.
    & CSI-Free
    & Extra Power \\
    \midrule

    \cite{10587082}
    & AN
    & $\times$
    & \checkmark
    & $\times$
    & $\times$ \\

    \cite{11602055,10605793,10804654}
    & BF
    & $\times$
    & \checkmark
    & $\times$
    & $\times$ \\

    \cite{11202391}
    & WD
    & $\times$
    & \checkmark
    & \checkmark
    & \checkmark \\

    \cite{zhang2026ambsentry}
    & AIoT
    & $\times$
    & \checkmark
    & \checkmark
    & \checkmark \\

    \midrule

    This work
    & ES
    & \checkmark
    & \checkmark
    & \checkmark
    & \checkmark \\

    \bottomrule[1.5pt]
    \end{tabular}

    \begin{tablenotes}
        \footnotesize
        \item AN: Artificial Noise;
        BF: Beamforming;
        WD: Waveform Design;
        ES: Environmental scatterers;
        Param. Prot.: Target Parameter Protection;
        CSI-Free: the warden-CSI-Free;
        AIoT: Ambient IoT.
        \checkmark: Supported;
        $\times$: Not Supported.
    \end{tablenotes}

    \end{threeparttable}%
    }
\end{table}

Overall, as summarized in Table~\ref{table_1}, existing sensing-security approaches remain limited. AN-assisted protection requires dedicated jamming power and often relies on the warden's CSI or spatial information, while secure beamforming similarly depends on accurate knowledge of the warden's channel or location. Sensing-secure waveform design can reduce such CSI dependence, but mainly protects target parameters rather than concealing the target's existence. A closely related approach, AmbSentry~\cite{zhang2026ambsentry}, exploits ambient IoT devices as cooperative jammers and ghost targets to degrade unauthorized sensing. However, it does not explicitly formulate target-existence covertness or require the warden's optimal detection performance to approach random guessing. In contrast, this work harnesses environmental scatterers as endogenous security resources and directly targets statistical indistinguishability between target-absent and target-present observations, without requiring the warden's CSI or dedicated jamming power.

\section{Endogenous Target Covertness}\label{sec:target_covertness}

To move beyond conventional sensing-security approaches that primarily suppress unauthorized sensing performance, we introduce endogenous target covertness as a new paradigm for protecting the physical existence of sensing targets. The key idea is to harness environmental scatterers as intrinsic security resources, thereby creating target-like uncertainty at the warden while preserving reliable sensing at the legitimate ISAC receiver.

\subsection{Target Covertness Definition}

Target covertness refers to concealing the physical existence of a target from unauthorized wardens while the legitimate sensing receiver can still detect the target. Specifically, the target detection of the warden can be formulated as a classic binary hypothesis testing problem: the warden must continuously monitor the radio frequency (RF) environment to decide whether the space contains only the baseline background ($\mathcal{H}_0$) or if a target is physically present ($\mathcal{H}_1$). The target becomes covert when the received signal distributions under both $\mathcal{H}_0$ and $\mathcal{H}_1$ are deliberately blurred, rendering them statistically indistinguishable from the warden's perspective.


\begin{definition}
A physical object is regarded as a covert target when an unauthorized warden cannot reliably distinguish between target absence and target presence, even under its optimal detection strategy. In this case, the warden's total detection error, jointly determined by false alarms and missed detections, approaches the level corresponding to random guessing. A stronger target covertness level, therefore, indicates greater statistical similarity between the observations with and without the protected target.
\end{definition}

\textbf{Physical Implication of the Definition:} This definition has a clear physical interpretation. When the warden's optimal detection performance approaches random guessing, the received sensing observations provide little reliable evidence about whether the protected target is physically present. In other words, the propagation environment has been shaped such that the target-induced echo is statistically obscured by the surrounding scattering responses, thereby concealing the target's physical existence from unauthorized sensing.

\begin{figure}[tp]
    \centering
    \includegraphics[width=0.9\linewidth]{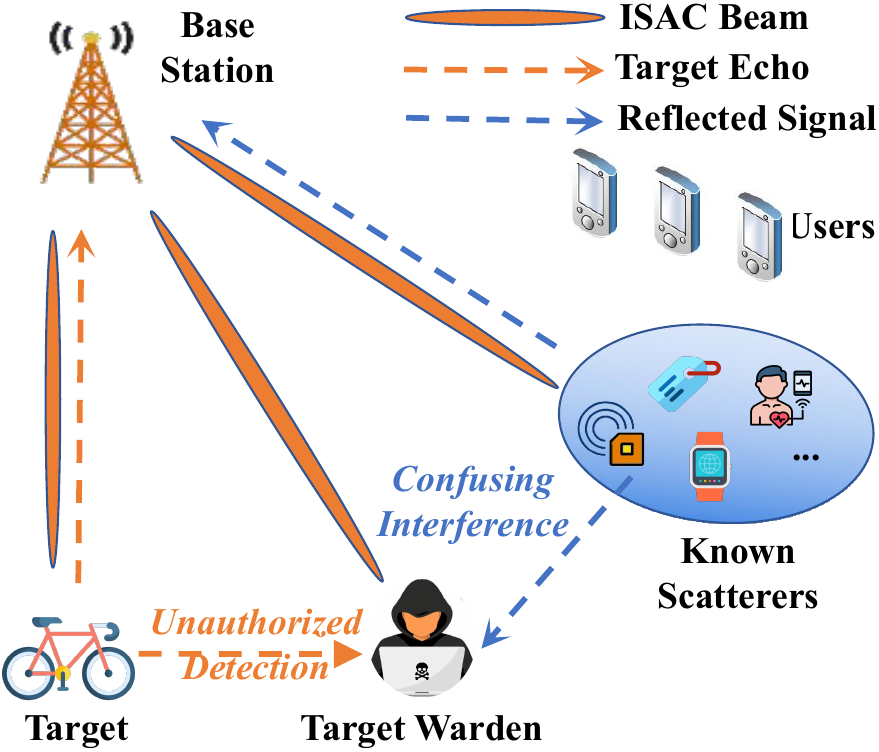}
    \caption{Illustration of environmental-scatterer-enabled endogenous target covertness in ISAC systems.}
    \label{syse}
\end{figure}

\subsection{The Proposed Endogenous Target Covertness Scheme}
\label{scheme}
The key idea of the proposed endogenous target covertness scheme is to deliberately exploit environmental scatterers located in the vicinity of the protected target as part of the sensing-security mechanism. As illustrated in Fig.~\ref{syse}, the base station (BS) forms an ISAC beam that simultaneously illuminates the true target and nearby scatterers. These scatterers reflect the incident waveform and consequently generate additional target-like echoes in the surrounding region. Since the target and the harnessed scatterers are spatially close to each other, their reflected signals may exhibit similar propagation delays and angular characteristics. As a result, from the perspective of an unauthorized sensing receiver, the true target echo is no longer observed as an isolated and easily distinguishable response, but is embedded within a cluster of spatially and dynamically correlated reflections.

This locally generated scattering ambiguity is particularly effective against a passive target-detector warden. The warden receives the superposition of echoes originating from both the true target and the surrounding scatterers, but lacks the prior information required to reliably determine which reflection corresponds to the protected target. Consequently, the warden should resolve two questions: 1) Is there a true target? 2) Which one is the true target? Those two questions result in false alarms and missed detections, respectively. Even when the warden has strong signal-processing capabilities, separating the true target response from nearby target-like reflections becomes significantly more difficult. More importantly, the resulting uncertainty affects not only the estimation of target parameters, but also the binary decision concerning target presence in the observed environment.

The legitimate ISAC receiver, in contrast, possesses asymmetric side information that enables it to distinguish the protected target from the intentionally harnessed scatterers. In practical deployments, the legitimate system can maintain prior knowledge of the protected target and the locations or reflection characteristics of participating scatterers through deployment information, previous sensing frames, device configuration, or sensing-assisted environmental mapping. Therefore, the scatterer-induced components are known or predictable from the legitimate receiver's perspective and can be reconstructed, filtered, or suppressed during sensing processing. After removing or mitigating these known reflections, the BS can recover the true target response and perform conventional target detection and parameter estimation. This creates a fundamental asymmetry between the legitimate receiver and the sensing warden: the same environmental reflections that appear as removable clutter to the legitimate BS act as unresolved target-like uncertainty to an unauthorized observer.


In this sense, the proposed scheme converts environmental scattering from a conventional sensing impairment into an endogenous security resource. Its protection capability is achieved by the propagation environment and the target's surrounding scatterers themselves, rather than relying exclusively on dedicated jamming signals or knowledge of the warden's instantaneous channel. By jointly exploiting local scattering similarity, legitimate-side prior knowledge, and controllable illumination power, the system is intended to conceal the existence of the protected target from unauthorized sensing while maintaining the required sensing performance at the legitimate receiver.

\subsection{Case Study: Environmental scatter-assisted Target Covertness Implementation}

In this section, we evaluate the proposed endogenous target covertness scheme from two perspectives: 1) whether environmental scatterers increase the warden's uncertainty in detecting the true target; and 2) whether the legitimate BS can preserve reliable sensing by exploiting prior knowledge of the target and surrounding scatterers. The simulation follows the model in Section~\ref{scheme}. For each realization, scatterer locations, large-scale path losses, and small-scale fading channels are generated according to a block-fading model. The BS transmits a common ISAC waveform that simultaneously illuminates the target and nearby scatterers. Since the BS knows the target and scatterer characteristics, the scatter-induced components can be reconstructed and suppressed, whereas the warden observes the superposition of the target echo and unresolved target-like reflections. We evaluate the warden's target detection error probability through Monte Carlo hypothesis testing and characterize the sensing performance of both receivers using received sensing SINR.

Following representative wireless sensing and ISAC settings \cite{skolnik2008radar,10587082,zhang2026ambsentry}, the BS, target, communication user, and warden are located at $(0,0)$~m, $(40,0)$~m, $(0,40)$~m, and $(40,20)$~m, respectively. Environmental scatterers are randomly distributed within $0.5$~m of the target. The BS and warden employ $16$ and $4$ antennas, respectively. The carrier frequency and bandwidth are $28$~GHz and $400$~MHz, with path-loss exponent $\alpha=2.2$ and reference gain $C_0=(\lambda/(4\pi d_0))^2$ at $d_0=1$~m. The thermal-noise density is $-174$~dBm/Hz, the BS power budget is $40$~dBm, the target scattering strength is $\sigma_T^2=1$, and the scatter reflection efficiency is $0.5$. A small residual scatter component is retained after legitimate-side cancellation to model practical imperfections, while the warden has no corresponding cancellation knowledge. We consider $K\in\{0,5,10,15,20\}$, where $K=0$ denotes conventional ISAC without scatterer assistance, and each sensing frame contains $L=64$ symbols. All results are averaged over independent channel and noise realizations.

\begin{figure}[tp]
    \centering
    \includegraphics[width=0.95\linewidth]{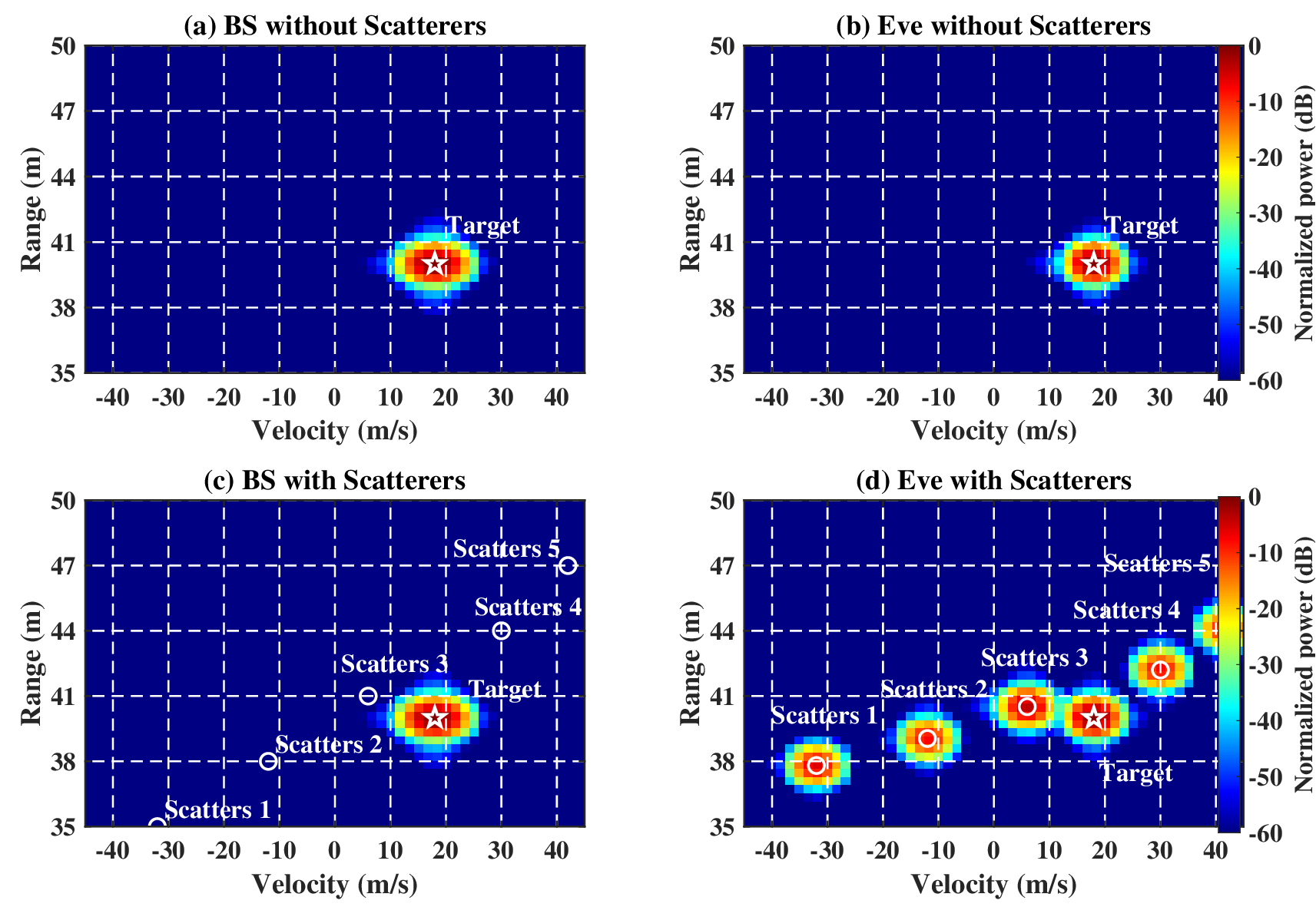}
\caption{Range--Doppler maps observed at the legitimate BS and the target warden with and without environmental-scatterer assistance.}
    \label{2}
\end{figure}
Fig.~\ref{2} illustrates the Range--Doppler responses of the proposed endogenous target covertness scheme at the legitimate BS and the warden. As shown in Fig.~\ref{2}(a) and Fig.~\ref{2}(b), without environmental scatterers, the true target produces a single dominant peak at both receivers, allowing the warden to readily infer its presence and motion. After introducing the scatterers, different observations arise. In Fig.~\ref{2}(c), the legitimate BS still maintains a clear and isolated target response because it knows the locations and reflection characteristics of the harnessed scatterers and can reconstruct and suppress their echoes before Range--Doppler processing.
In contrast, Fig.~\ref{2}(d) shows multiple target-like responses at the warden. Lacking the side information required to identify or cancel the scatter-induced echoes, the warden observes the true target embedded among several plausible reflections with comparable power levels. Consequently, it becomes difficult to reliably associate a Range--Doppler peak with the true target, increasing the ambiguity in determining the target's presence and sensing parameters. The comparison between Fig.~\ref{2}(c) and Fig.~\ref{2}(d) demonstrates the key principle of endogenous target covertness: environmental scattering can obscure unauthorized sensing while preserving reliable target sensing at the legitimate BS.

\begin{figure}[tp]
    \centering
    \includegraphics[width=0.9\linewidth]{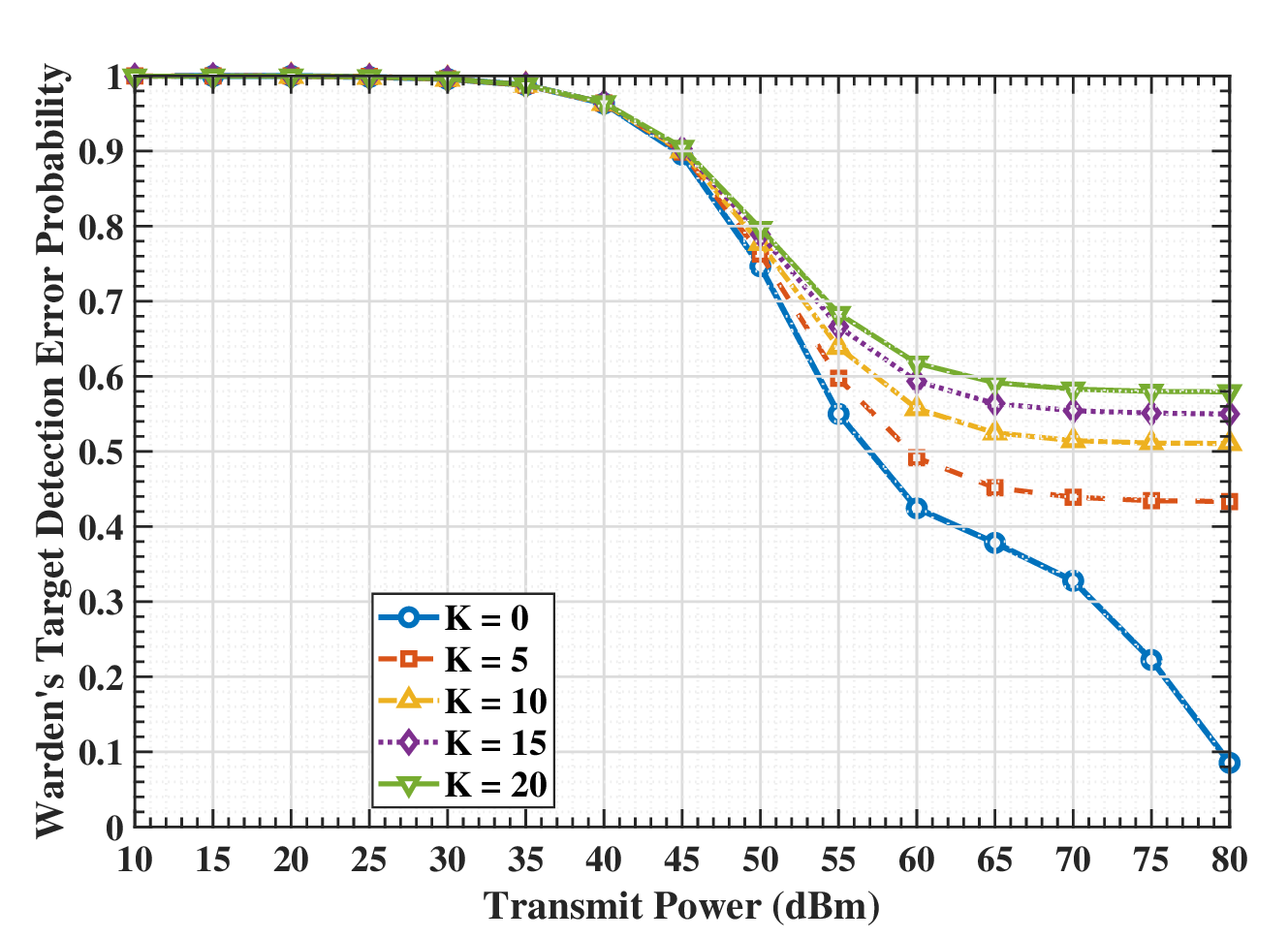}
\caption{Warden's minimum total target detection error probability versus the BS transmit power for different numbers of environmental scatterers.}
    \label{0}
\end{figure}

Fig.~\ref{0} illustrates the warden's target detection error probability versus the BS transmit power for different numbers of environmental scatterers. For the conventional ISAC baseline without scatter assistance ($K=0$), the detection error probability rapidly decreases with transmit power, since the stronger target echo facilitates reliable target detection. In contrast, with environmental scatterers ($K>0$), the detection error probability decreases initially but approaches a nonzero error floor at high transmit power. This occurs because increasing the transmit power simultaneously strengthens both the true target echo and the scatter-induced target-like reflections. Once receiver noise becomes negligible, the warden's detection performance is therefore mainly limited by the statistical distinguishability between the target and surrounding scattering responses rather than the absolute received power. Moreover, increasing the number of scatterers raises the error floor by introducing stronger target-like uncertainty. These results demonstrate that environmental scatterers provide a persistent covertness gain that cannot be removed simply by increasing the received sensing power.
\begin{figure}[tp]
    \centering
    \includegraphics[width=0.9\linewidth]{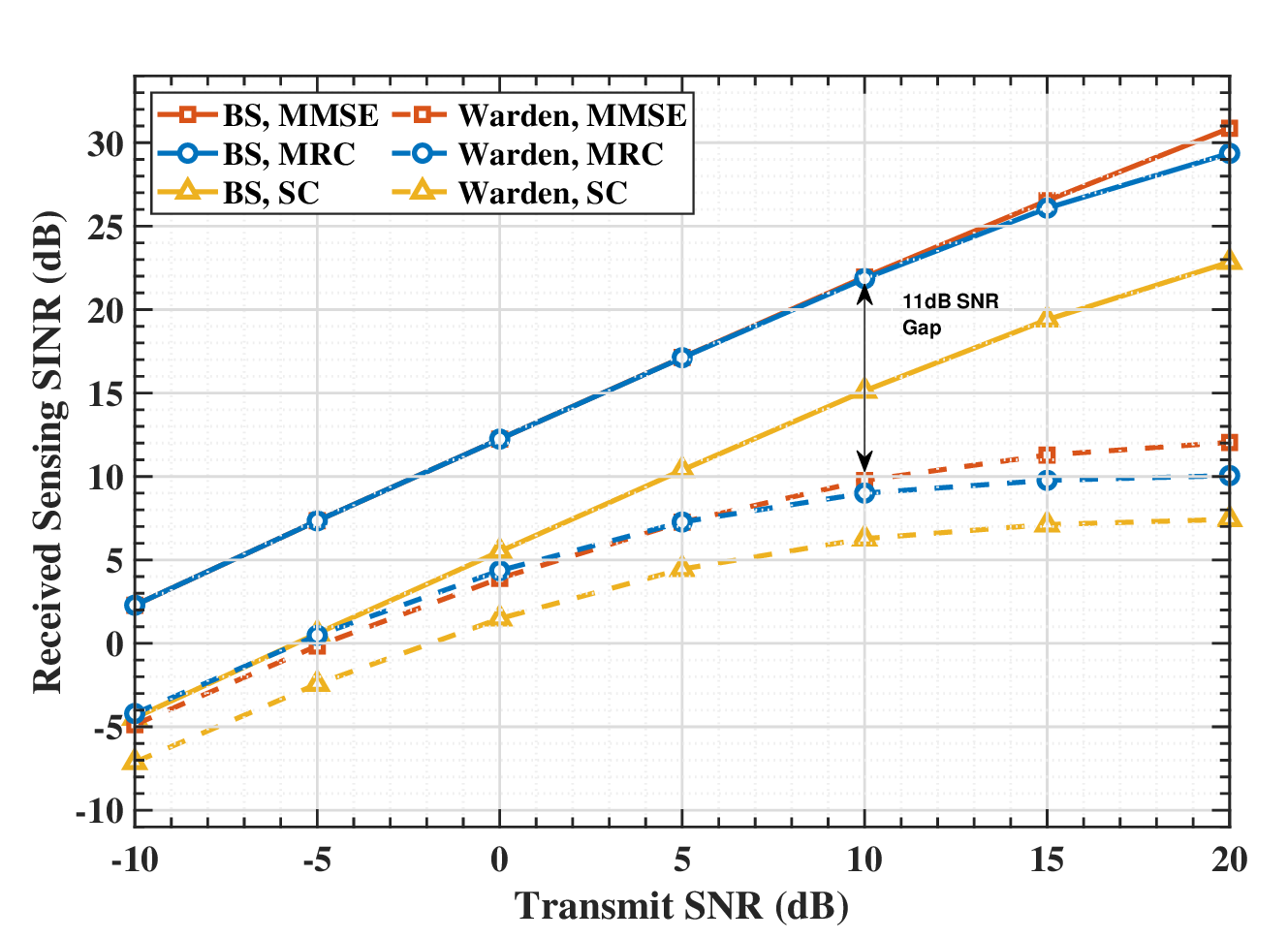}
\caption{Sensing-SINR comparison between the legitimate BS and the target warden under different linear combining schemes with $K=10$ environmental scatterers.}
    \label{1}
\end{figure}

Fig.~\ref{1} compares the received sensing SINR at the legitimate BS and the warden under different linear combining strategies. The BS consistently achieves a much higher sensing SINR because it can exploit prior knowledge of the environmental scatterers to suppress most scatter-induced reflections, whereas the warden observes the target echo together with unresolved target-like interference. At a transmit SNR of $10$~dB, an approximately $11$~dB SINR gap is observed, demonstrating the sensing advantage enabled by the proposed endogenous covertness mechanism.
This advantage persists across different receive-processing strategies. At the BS, advanced spatial processing further mitigates residual scatter interference, while the warden's SINR gradually saturates at high transmit SNR because its performance becomes interference-limited. These results confirm that the proposed scheme preserves reliable legitimate sensing while degrading the warden's ability to extract the true target response.

\section{Current Issues and Future Directions} \label{sec:future_directions}
Although target covertness offers a promising mechanism for protecting physical targets against unauthorized sensing, its practical realization still faces challenges in resource utilization and fundamental performance characterization. This section highlights promising research directions toward more controllable, efficient, and theoretically grounded target-covertness mechanisms for future 6G ISAC systems.

\subsection{Multi-Dimensional Resource Optimization}
\textbf{\textit{\textbf{Current Issue:}}} Achieving optimal target covertness requires careful coordination of network resources across multiple domains. Currently, dynamically allocating resources, such as transmit power, beamforming vectors, and frequency subcarriers, between the target and the scatterers is a highly complex, non-convex optimization problem. For instance, while widening the sensing beam to illuminate more scatterers increases target warden confusion, it inherently dilutes the energy hitting the target, potentially compromising legitimate sensing.

\textbf{\textit{Future Direction Suggestion:}} Future research should develop low-latency, robust optimization algorithms capable of adapting to rapidly changing environments and highly mobile targets. Leveraging artificial intelligence, particularly deep reinforcement learning (DRL), offers a promising avenue to achieve real-time, adaptive resource allocation. These AI-driven approaches can intelligently balance beamforming and power distribution to adapt to intelligent wardens in highly dynamic channel conditions.

\subsection{Integration of Diverse Controllable Scatterers}
\textbf{\textit{\textbf{Current Issue:}}} Seamlessly integrating massive numbers of controllable devices into ISAC networks raises critical practical hurdles. Specifically, managing heterogeneous devices introduces massive channel estimation overhead, tight synchronization requirements between the base station and the scatterers, and challenges related to mitigating hardware impairments, all of which currently limit the scalability of endogenous covertness.

\textbf{\textit{Future Direction Suggestion:}} A significant future direction involves transitioning from simple ambient reflectors to orchestrating a fully programmable physical environment using advanced RISs and ambient backscatter devices. By densely deploying RISs and BDs, the ISAC network can proactively steer signal reflections, manipulate phase shifts, and introduce time-varying modulations to synthesize highly convincing false Doppler profiles and spatial clutter, thereby providing much stronger and precisely controllable target concealment. In addition, time-modulated corner reflectors (TMCRs) represent another promising class of controllable passive scatterers, as their time-varying reflection characteristics can deliberately reshape the Doppler and imaging signatures observed by unauthorized sensing receivers, thereby creating additional target-like ambiguity without requiring active RF transmission \cite{10955353}.

\subsection{Information-Theoretic Bounds on Target Covertness}
\textbf{\textit{\textbf{Current Issue:}}} To mature from an empirical technique into a rigorous scientific framework, the field currently lacks the foundational theoretical limits for sensing concealment. Just as Shannon capacity defines the absolute limit of communication, the absence of theoretical upper bounds for target covertness makes it difficult to benchmark the absolute efficiency and limits of practical secure ISAC algorithms.

\textbf{\textit{Future Direction Suggestion:}} There is an urgent need to establish the information-theoretic foundations of target covertness. Future studies should focus on deriving rigorous mathematical limits, such as utilizing Kullback-Leibler (KL) divergence or mutual information metrics, to quantify the maximum achievable covertness capacity of an ISAC system given a specific number of antennas, available bandwidth, and scatterer density, independent of the specific detection techniques employed by the target warden.

\subsection{Trade-offs and Pareto Boundaries in Secure ISAC}
\textbf{\textit{\textbf{Current Issue:}}} At its core, the design of a secure ISAC system is governed by a fundamental, multi-dimensional tradeoff among communication throughput, legitimate sensing accuracy, and target covertness. Allocating degrees of freedom or power to confuse the target warden inevitably consumes valuable resources that could otherwise be used to enhance data rates or improve sensing resolution.

\textbf{\textit{Future Direction Suggestion:}} A crucial future direction is the exhaustive mathematical exploration of this three-way trade-off relationship. Researchers should strive to analytically map the Pareto boundary where no single metric can be improved without degrading another. Understanding and modeling this precise Pareto front will empower network operators with flexible design guidelines, allowing them to dynamically shift the system's operating point to satisfy the specific service level agreements (SLAs) of diverse 6G applications.

\subsection{Resilience Against Active and Complex Attacks}
\textbf{\textit{\textbf{Current Issue:}}} Most existing sensing security frameworks fundamentally assume a strictly passive target warden who merely intercepts ambient signals. However, sophisticated adversaries in future 6G networks may deploy active attack capabilities. These include sensing jamming to deliberately blind the legitimate perception system, sensing spoofing to inject phantom targets and cause misjudgments, and active eavesdropping where the attacker emits their own probing waveforms to forcefully illuminate the target and bypass passive concealment strategies.

\textbf{\textit{Future Direction Suggestion:}} A highly compelling research frontier is expanding the role of environmental scatterers, BDs, and RISs from passive camouflage to active physical-layer defense. Future studies should explore how these programmable elements can be repurposed to counter active threats. For instance, RIS could be configured to dynamically steer deep spatial nulls toward jamming sources, synthesize attractive ghost targets to confuse sensing spoofers, or actively scramble the echoes generated by an active target warden's probing signals, thereby establishing a comprehensive, multidimensional sensing defense shield.



\section{Conclusion} \label{sec:conclusion}
As 6G networks evolve to seamlessly bridge the digital and physical worlds, securing the sensing functionality of ISAC systems becomes as critical as securing the communication data itself. This article has highlighted a fundamental vulnerability in current sensing security frameworks: the inability to hide the physical existence of a target from unauthorized wardens. To overcome this, we introduced the novel paradigm of endogenous target covertness. By shifting the perspective of environmental scatterers from unwanted propagation clutter to highly valuable security assets, we demonstrated that it is physically possible to camouflage a target's presence within a dynamic radio frequency background. This approach forces the target warden's detection capability toward random guessing while strictly preserving the high-precision perception of legitimate ISAC nodes. Furthermore, we identified open issues and suggested future research directions. Ultimately, embedding target covertness directly into the physical layer will be indispensable for unlocking the full, secure potential of ubiquitous sensing in next-generation wireless networks.

\bibliographystyle{IEEEtran}
\bibliography{reference}

\end{document}